\documentclass[twocolumn]{aastex61}
\usepackage{silence}
\usepackage{graphicx}
\usepackage[caption=false]{subfig}
\usepackage{booktabs}
\usepackage{stmaryrd}

\usepackage{natbib}
\setcitestyle{authoryear,open={(},close={)}}

\usepackage{float}
\usepackage{color,soul}
\usepackage{wrapfig}
\usepackage{bm}
\usepackage{amsmath}

\usepackage{fancyhdr}
\usepackage{wasysym}

\newcommand{\cii}{[\ion{C}{2}]}
\newcommand{\oi}{[\ion{O}{1}]}

\newcommand{\lir}{L$_{\mathrm{IR}}$}
\newcommand{\lcii}{L$_{[\mbox{\footnotesize \ion{C}{2}}]}$}
\newcommand{\loi}{L$_{[\mbox{\footnotesize \ion{O}{1}}]}$}
\newcommand{\lsiii}{L$_{[\mbox{\footnotesize \ion{Si}{2}}]}$}

\newcommand{\loglir}{$\log\mathrm{L}_{\mathrm{IR}}/\mathrm{L}_\odot$}
\newcommand{\sigmaIR}{$\Sigma_{\mathrm{IR}}$}
\newcommand{\logSigmaIR}{$\log\Sigma_{\mathrm{IR}}/\mathrm{[L_\odot\,kpc^{-2}]}$}

\newcommand{\lpah}{L$_{\mathrm{\tiny PAH}}$}

\newcommand{\ephot}{$\epsilon_{\tiny\mathrm{PE}}$}
\newcommand{\fagn}{$f_{\tiny\mathrm{AGN,MIR}}$}
\newcommand{\fpdr}{$f_{\tiny\mathrm{PDR}}$}

\newcommand{\siii}{[\ion{Si}{2}]}
\newcommand{\Go}{G$_0$}
\newcommand{\nH}{n$_{\tiny\mathrm{H}}$}
\newcommand{\LelevenThree}{$\mathrm{L_{11.3\,\mu m}/L_{3.3\,\mu m}}$}
\newcommand{\LsixSeven}{$\mathrm{L_{6.2\,\mu m}/L_{7.7\,\mu m}}$}
\newcommand{\LsevenSix}{$\mathrm{L_{7.7\,\mu m}/L_{6.2\,\mu m}}$}
\newcommand{\LelevenSeven}{$\mathrm{L_{11.3\,\mu m}/L_{7.7\,\mu m}}$}

\begin{document}

\title{Regulating star formation in nearby dusty galaxies: Low photoelectric efficiencies in the most compact systems}
\author{J. McKinney} 
\affiliation{Department of Astronomy, University of Massachusetts, Amherst, MA 01003, USA.}
\affiliation{Infrared Processing and Analysis Center, MC 314-6, Caltech, 1200 E. California Blvd., Pasadena, CA 91125, USA.}
\author{L. Armus}
\affiliation{Infrared Processing and Analysis Center, MC 314-6, Caltech, 1200 E. California Blvd., Pasadena, CA 91125, USA.}
\author{A. Pope}
\affiliation{Department of Astronomy, University of Massachusetts, Amherst, MA 01003, USA.}
\author{T. D\'iaz-Santos}
\affiliation{N\'ucleo de Astronom\'ia de la Facultad de Ingenier\'ia y Ciencias, Universidad Diego Portales, Av. Ej\'ercito Libertador 441, Santiago, Chile}
\affiliation{Chinese Academy of Sciences South America Center for Astronomy, National Astronomical Observatories, CAS, Beijing 100101, China}
\affiliation{Institute of Astrophysics, Foundation for Research and Technology-Hellas, GR-71110, Heraklion, Greece}
\author{V. Charmandaris}
\affiliation{Institute of Astrophysics, Foundation for Research and Technology-Hellas, GR-71110, Heraklion, Greece}
\affiliation{Department of Physics, University of Crete, GR-71003, Heraklion, Greece}
\author{H. Inami}
\affiliation{Hiroshima Astrophysical Science Center, Hiroshima University, 1-3-1 Kagamiyama, Higashi-Hiroshima, Hiroshima 739-8526, Japan}
\author{Yiqing Song}
\affiliation{Astronomy Department, University of Virginia, 530 McCormick Road, Charlottesville, VA 22904, USA}
\author{A.S. Evans}
\affiliation{National Radio Astronomy Observatory, 520 Edgemont Road, Charlottesville, VA 22903, USA}
\affiliation{Astronomy Department, University of Virginia, 530 McCormick Road, Charlottesville, VA 22904, USA}

\begin{abstract}
Star formation in galaxies is regulated by the heating and cooling in the interstellar medium. In particular, the processing of molecular gas into stars will depend strongly on the ratio of gas heating to gas cooling in the neutral gas around sites of recent star-formation. In this work, we combine mid-infrared (mid-IR) observations of Polycyclic Aromatic Hydrocarbons (PAHs), the dominant heating mechanism of gas in the interstellar medium (ISM), with \cii, \oi, and \siii\ fine-structure emission, the strongest cooling channels in dense, neutral gas. The ratio of IR cooling line emission to PAH emission measures the photoelectric efficiency, a property of the ISM which dictates how much energy carried by ultraviolet photons gets transferred into the gas. We find that star-forming, IR luminous galaxies in the Great Observatories All-Sky LIRG Survey (GOALS) with high IR surface densities have low photoelectric efficiencies. These systems also have, on average, higher ratios of radiation field strength to gas densities, and larger average dust grain size distributions. The data support a scenario in which the most compact galaxies have more young star-forming regions per unit area, which exhibit less efficient gas heating. These conditions may be more common at high$-z$, and may help explain the higher star-formation rates at cosmic noon. We make predictions on how this can be investigated with \textit{JWST}.
\end{abstract}

\section{Introduction}

Star-formation in a galaxy is dependent on processes which remove energy allowing gas to cool. Only the coldest phases will collapse under self-gravity to form stars, and so characterizing the mechanisms by which gas heats and cools is critical to our understanding of star-formation, and galaxy evolution. The cold gas associated with star-formation emits strongly in low energy atomic and molecular transitions, bright at infrared (IR) wavelengths. 

Studies with the \textit{Infrared Space Observatory (ISO)} were key in unveiling the $5-200\mu m$ wavelength regime of the electro-magnetic spectrum of galaxies, which is rich in strong atomic and molecular emission lines that trace the ISM \citep{Malhotra1997,Malhotra2001,Dale2000}. Amongst the brightest of these features are \oi\ $63\,\mu m$, and \cii\ $157.7\,\mu m$ which can contain $0.1-1\%$ of total infrared luminosities \citep{Malhotra2001,Stacey2010,DiazSantos2013,Brisbin2015,Ibar2015}. The \textit{Herschel Space Telescope} was used to significantly increase the number of galaxies detected in the far-IR lines (e.g., \citealt{Sturm2000,vanderWel2014,DiazSantos2011}), and studies with the \textit{Spitzer Space Telescope} added other prominent cooling lines such as \siii\ $34.8\,\mu m$, as well as Polycyclic Aromatic Hydrocarbon (PAH) vibrational lines between $3-17\,\mu$m that trace photoelectric heating. 

\cii, \oi, and \siii\ are well-established as strong coolants in Milky Way photodissociation regions (PDRs), the transition zones between \ion{H}{2} regions and the cold-neutral-medium (e.g., \citealt{Wolfire1990,Wolfire2003}). As the boundary layer between warm and cold gas, PDRs are an excellent place to study the processes of heating and cooling, as the relative balance of energy gains and losses can impact future star-formation. Emission from small grains, PAHs, provide an excellent tracer of the photoelectric heating of gas in PDRs \citep{Galliano2008,Tielens2008}. Photoelectrons ejected from PAH grains share energy with H and H$_2$, which go on to collisionally excite fine-structure transitions of C$^+$, O, and Si$^+$. Although \cii\ and \oi\ emission are the dominant cooling channels in more normal PDRs (e.g., \citealt{Tielens1985,Kaufman1999}), \siii\ can also act as a significant coolant at high interstellar radiation field densities \citep{Kaufman2006}. Thus, the cycle of gas heating and cooling in PDRs is traced by \cii, \oi, \siii, and PAH emission, which collectively make for a powerful diagnostic of the ISM in extragalactic sources (e.g., \citealt{Malhotra2001,Helou2001,Croxall2012,Beirao2012,Sutter2019}. 

The deficit of \cii\ emission per unit \lir\ in warmer (higher dust temperatures; T$_\mathrm{dust}$), and more compact Luminous IR Galaxies (LIRGs: \loglir$\,=11-12$, ULIRGs: \loglir$\,>12$) as compared to less extreme galaxies is a well-studied phenomenon (e.g., \citealt{Malhotra1997,Malhotra2001,Luhman1998,Luhman2003,Stacey2010,HerreraCamus2015,DiazSantos2013,DiazSantos2014,DiazSantos2017,Ibar2015,Smith2017}). Moreover, PAH emission and other far-IR fine-structure lines key to PDR heating and cooling show similar deficits (e.g., \citealt{Brauher2008,Wu2010,GC2011,Pope2013,Stierwalt2014,DeLooze2014,Cormier2015,DiazSantos2017}). The IR emission line deficits can be due to changes in ISM densities and the strength of the radiation field impinging upon PDR surfaces (e.g., \citealt{DiazSantos2017}), and/or opacity effects, thermal line saturation, and evolution in dust grain properties \citep{Luhman1998,Malhotra2001,Helou2001,MunozOh2016,Smith2017}; however, the deficits largely disappear in normal star-forming galaxies when the line emission is normalized by total PAH emission \citep{Helou2001,Croxall2012,Sutter2019}. This reinforces the link between the ejection of photoelectrons from PAH grains and the collisional excitation of \cii\ and other fine-structure lines, and suggests that the heating and cooling processes within PDRs behave similarly for normal star-forming conditions over a range of T$_{\mathrm{dust}}$ and \lir. Interestingly, \cite{McKinney2020} found lower \cii/PAH emission at $z\sim2$ than $z\sim0$, suggesting evolution in the heating and cooling balance with redshift. However, the cooling/heating ratio has yet to be fully characterized in $z\sim0$ starbursts, which represents a key step for future studies of the early Universe. 

In this work, we explore the range of the ratio between fine-structure line cooling to photoelectric heating in local dusty, star-forming galaxies by comparing \cii, \oi, and \siii\ emission to PAH emission in the \textit{Great Observatories All Sky LIRG Survey} (GOALS; \citealt{Armus2009}). GOALS is comprised of 244 galaxy nuclei within 202 LIRGs and ULIRGs spanning a range in merger stage and morphology. The range in IR luminosities and stellar masses of galaxies in GOALS makes the sample a bridge between normal star-forming galaxies and the most extreme, compact starbursts that host atypical PDR conditions exposed to stronger radiation fields \citep{DiazSantos2017}. Whether or not the gas cooling/heating properties are fundamentally different in such extreme environments remains an open question which this work aims to address. 

GOALS has been the subject of extensive study with observations spanning the electromagnetic spectrum\footnote{For a complete list, visit \url{http://goals.ipac.caltech.edu/publications.html}.}, including mid- and far-IR measurements of PDR cooling lines and PAHs, the ratio of which is an empirical measure of the photoelectric efficiency, \ephot: the fraction of energy, in UV photons, emitted by hot stars that is absorbed by small dust grains and transferred into the neutral PDR gas via the photoelectric effect (e.g., \citealt{Gerola1976,deJong1977}). There is evidence that \ephot\ is a constant amongst normal star-forming galaxies \citep{Helou2001,Sutter2019}; however, this has yet to be tested in the starburst regime typical in luminous and ultra-luminous infrared galaxies powered by star formation.. 

The paper is organized as follows: We review the multi-wavelength observations in Section \ref{sec:obs}, and discuss the sample selection. Section \ref{sec:DataAnalysis} summarizes important derived quantities of GOALS galaxies presented in other works, as well as key analysis steps we take in this paper to appropriately combine the multi-wavelength data. Our results are presented in Section \ref{sec:results}, which we discuss in Section \ref{sec:discussion} in the context of local star-formation and trends in galaxy evolution. Section \ref{sec:summary} summarizes the paper. 

\section{Sample Selection and Data \label{sec:obs}}
This work focuses on data from \textit{Spitzer}/IRS \citep{DiazSantos2010,DiazSantos2011,Petric2011,Stierwalt2013,Stierwalt2014,Inami2013}, \textit{Spitzer}/IRAC (Mazzarella et al. 2020, in prep.), \textit{Herschel} PACS and SPIRE \citep{Zhao2013,DiazSantos2013,DiazSantos2014,DiazSantos2017}, and AKARI/IRC \citep{Inami2018}.

To measure the ratio of cooling and heating in star-forming gas, we select from the 244 galaxies in GOALS all sources with [\ion{Ne}{5}]$_{14.3\mu m}$ / [\ion{Ne}{2}]$_{12.8\mu m} <0.5$ and/or [\ion{O}{4}]$_{25.9\mu m}$ / [\ion{Ne}{2}]$_{12.8\mu m} <1$ in oder to exclude from the sample those galaxies with significant heating from a central AGN \citep{Inami2013,Petric2011}. Sources with low PAH equivalent widths (EW) but no high-ionization emission lines are not removed from the sample to avoid biasing the results concerning grain properties such as size and ionization. As a result, 16 strong AGN systems are removed from consideration. Heavily dust obscured AGN could still be lurking in the sample, and we use mid-IR AGN fractions (\fagn) derived in \cite{DiazSantos2017} to identify sources with excess warm dust in the remaining sample, which we show in Figures to concentrate on the star-forming properties of the sample galaxies. 

To fully explore the properties of interstellar PAH grains, we use AKARI $2.5-5\,\mu m$ spectra of 145 GOALS (U)LIRGs presented in \cite{Inami2018}, in which the 3.3$\,\mu m$ PAH feature was detected for 133 targets. This sub-sample spans the full range of \lir\ and luminosity distance, and is representative of the range in star-formation properties of GOALS. 

\section{Analysis\label{sec:DataAnalysis}}
In addition to measured line fluxes, this work makes use of a number of quantities from \textit{Herschel}/PACS observations derived in \cite{DiazSantos2017}: We subtract out the ionized component of \cii\ emission using the neutral \cii\ PDR fractions (\fpdr), estimated with [\ion{N}{2}]$_{122}$ and [\ion{N}{2}]$_{205}$ available for 54\% of the sample, and far-IR colors $S_{\nu,63\mu m}/S_{\nu,158\mu m}$ otherwise, as described in \cite{DiazSantos2017}. We use IR surface densities from \cite{DiazSantos2017} which are calculated from the effective area measured at $70\,\mu m$, and total IR luminosities from \cite{Armus2009}. We also use measurements of the average intensity of the UV interstellar radiation field impinging onto the surface of PDRs, G, measured in local units (\Go$\,=1.6\times10^{-3}$ erg s$^{-1}$ cm$^{-2}$, \citealt{Habing1968}), and neutral gas volume densities, \nH, derived using the \cite{Kaufman1999,Kaufman2006} PDR models through \texttt{PDR TOOLKIT} (PDRT; \citealt{PDRT}) which depend principally on the galaxy-integrated \cii\ and \oi\ line fluxes, as well as \lir. 

\begin{figure}
    \centering
    \includegraphics[width=.5\textwidth]{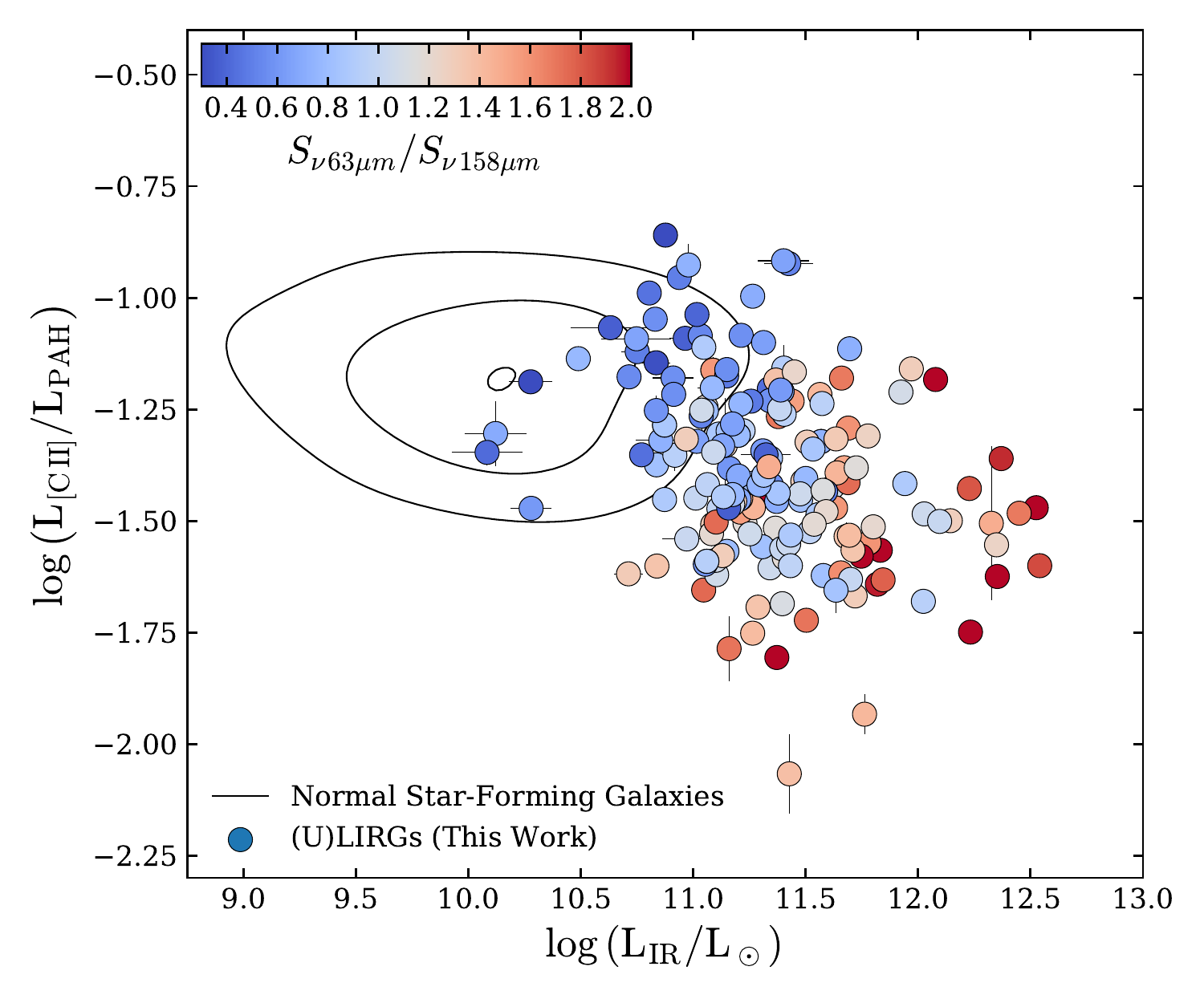}
    \caption{The ratio of \cii\ emission to the sum of 6.2, 7.7, 8.6, 11.3 and 17$\,\mu m$ PAH emission in the nuclear regions of GOALS star-forming galaxies as a function of total IR luminosity. The ionized gas contribution to \lcii\ has not been subtracted out. Each data point is colored by the far-IR color measured with the ratio of continuum at 63$\,\mu m$ and $158\,\mu m$. Black contours contain 33\%, 66\%, and 95\% of normal star-forming galaxies measured by \textit{ISO} \citep{Helou2001,Malhotra2001,Dale2000}. Note that we have scaled the ISOCAM PAH measurements by a factor of 42\% to estimate \lpah\ from the 6.75$\,\mu m$ broadband photometry, correcting for the underlying continuum and longer wavelength features. The correlation between \cii/PAH and \lir\ in local LIRGs and ULIRGs is weak, with a great deal of scatter, but warmer objects tend to have lower \cii/PAH ratios compared to normal star-forming galaxies and cooler LIRGs.}
    \label{fig:cii_pah_vs_lir}
\end{figure}

\subsection{PAH Properties\label{sec:pahprops}}
The PAH line fluxes used in our analysis are reported in \cite{Stierwalt2013,Stierwalt2014}, and \cite{Inami2018}. We take the PAH luminosity, \lpah, to be the sum of features measured between $\sim5-18\,\mu m$ by \textit{Spitzer} IRS. Specifically, \lpah\ includes the $6.2$, $7.7$, $8.6$, $11.3$, and $17\,\mu m$ PAH lines and all sub-features therein, as measured by \texttt{CAFE} which fits the line fluxes, continuum and the extinction simultaneously \citep{Marshall2007}. On average, these five PAH lines account for $76\%\pm9\%$ of the total PAH luminosity measured by both the Short-Low (SL, $5.5-14.5\,\mu m$) and Long-Low (LL, $14-38\,\mu m$) slits in GOALS galaxies (\citealt{Stierwalt2014}). Because the SL and LL slits have different widths, tied to the changing PSF with wavelength, some highly resolved galaxies show a small jump between their SL and LL IRS spectra. This is discussed fully in \cite{Stierwalt2013}, and we combine the SL and LL data using their scale factors derived from the wavelength overlap between the two slits. \lpah\ is normalized to the SL slit. The 3.3$\,\mu m$ PAH luminosities ($\mathrm{L_{3.3}}$) are also matched to the aperture extraction of IRS/SL data (see \citealt{Inami2018} Section 3); however, $\mathrm{L_{3.3}}$ represents a small component of \lpah\ ($5\%\pm1.5\%$), and is therefore only used to investigate the properties of PAH line ratios. In general, the fraction of total PAH power measured through the SL slit compared to the LL slit does not correlate with galaxy distance, or other quantities we compare against in this work, namely \sigmaIR. The sum of PAH features between 6 and 18$\,\mu m$ traces the bulk of the total power of PAH emission in GOALS. 

From the sub-sample of GOALS galaxies with AKARI observations, 50 objects were selected as test-cases for \texttt{CAFE-NG} (Advanced \texttt{CAFE}; Bonfini et. al., in prep., \citealt{Marshall2007}), which simultaneously fits continuum and spectral features in the $0.6-35\,\mu m$ range. These galaxies have full spectral model fits to the AKARI+Spitzer data, which explicitly includes silicate absorption and 3.05$\,\mu m$ ice feature absorption which are fit independently. Joint fits to IR continuum and line measurements are important to get accurate line fluxes e.g., \citealt{Smith2007,Lai2020}), and we use these new \texttt{CAFE} fits to estimate extinction-corrected 3.3$\,\mu m$ luminosities, $\mathrm{L_{3.3}}$, for the full AKARI sample. In general, the measured PAH intensity depends on how the local continuum is estimated. \cite{Inami2018} adopt a spline technique ($\mathrm{L_{3.3,spline}}$), which is known to under-estimate the extinction-corrected flux of PAH features between $6-17\,\mu$m by $30-60\%$ \citep{Smith2007}. From careful fits to the full PAH features (Bonfini et al., in prep.) we find that $\mathrm{L_{3.3,spline}/L_{3.3}}=0.29\pm0.07$. For the sample here we use this conversion and the results from \cite{Inami2018} to estimate  $\mathrm{L_{3.3}}$ for each source before comparing to the other PAH features. 

\subsection{Aperture Matching\label{sec:aptrs}}
Galaxies in GOALS vary from extended sources spatially resolved by \textit{Spitzer} to unresolved point-like sources, which motivates a careful approach to aperture matching when comparing \textit{Sptizer} and \textit{Herschel} measurements. The effective area of the \textit{Spitzer}/IRS SL slit is $3''.7\times9''.5$ \citep{Stierwalt2013}. To best match this area, we begin by using \cii\ and \oi\ line fluxes measured through the central spaxel of the PACS/IFU with dimensions of $\sim9''.4\times9''.4$ (See Sect.\,3.2 in \citealt{DiazSantos2017}). Next, we project the effective SL slit and PACS central-spaxel aperture onto the \textit{Spitzer}/IRAC $8\,\mu m$ images, assuming that it traces the co-spatial PAH and far-IR line emission from star-forming regions in LIRGs, which is reasonable for the $2-7$ kpc-scale observations in this work \citep{Peeters2004,AlonsoHerrero2010,PereiraSantaella2010,Salgado2016,Hughes2016}. \cite{Smith2007} show the fractional power of PAHs to the IRAC 8$\,\mu m$ band to have a median value of $70\%$ in normal star-forming galaxies, which increases only marginally when a more careful subtraction of stellar continuum is done using the IRAC $3-5\,\mu m$ channels (e.g., \citealt{Helou2004}). We use the ratio of the $8\,\mu m$ flux through each aperture to derive a correction, which is then used to scale down the PACS measurements to match the IRS aperture. The median of aperture corrections derived in this manner is 0.62 with upper and lower quartiles of 0.69 and 0.55 respectively. We note that aperture matching does not introduce any bias in the slope of trends we explore in this paper. 

The \siii\ 34.8$\,\mu m$ emission was measured in the \textit{Spitzer}/IRS LL slit, using an effective extraction aperture of $\sim10''.6\times36''.6$ \citep{Stierwalt2013}. To match measurements made through the IRS/SL slit, we scale the \siii\ line flux down to the normalization of the SL spectrum using the same multiplicative factors applied to the 17$\,\mu m$ PAH fluxes \citep{Stierwalt2013}. As noted in \cite{Stierwalt2014}, 15 LIRGs in GOALS have scale factors $>2$ due to extended structure in the LL slit not captured by the SL slit. To avoid complications, we do not include these sources in our analysis. In summary, we have matched the apertures between \textit{Herschel}/PACS and \textit{Spitzer}/IRS, and the measurements reported here are for the central regions ($\sim2-7$ kpc) of GOALS systems unless explicitly stated otherwise. 


\begin{figure*}[ht]
    \centering
    \includegraphics[width=\textwidth]{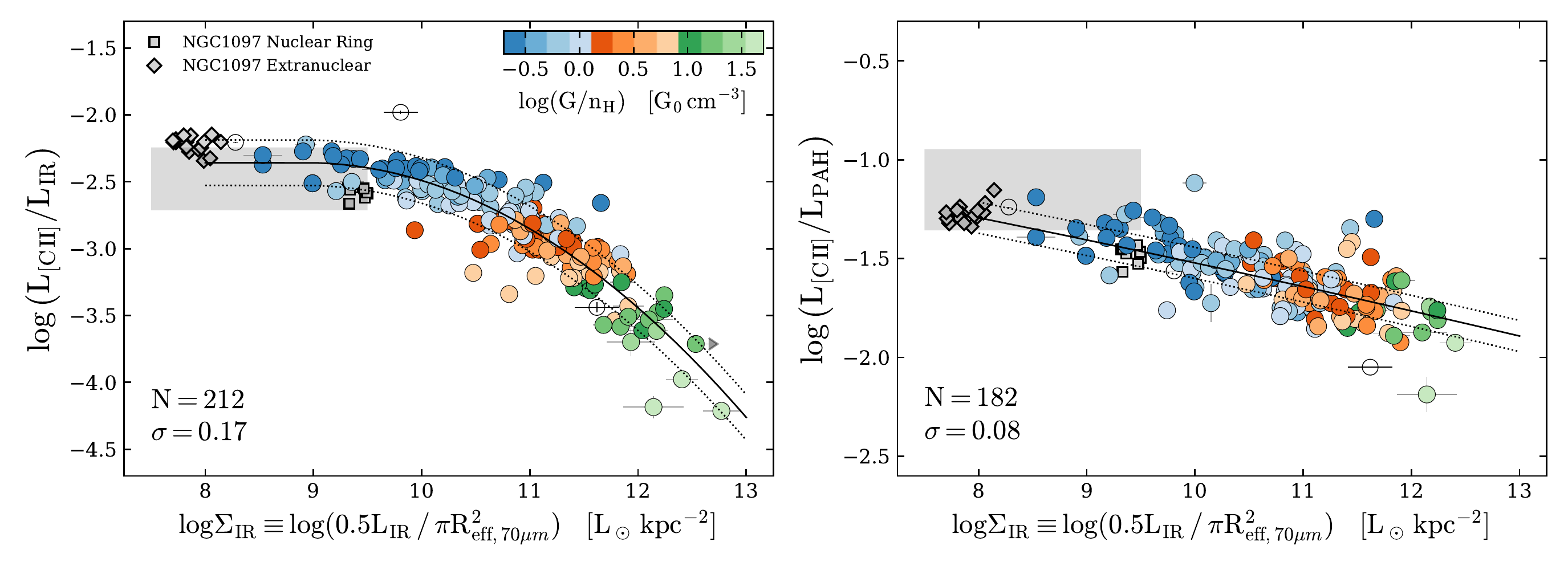}
     \caption{(\textit{Left}) The \cii\ deficit in GOALS as presented by \cite{DiazSantos2017}, calculated with galaxy-integrated measurements of \lcii\ and \lir. The data are colored by their average ratio of UV radiation field strength to neutral hydrogen density in PDRs (G/$\mathrm{n_H}$). The solid black line indicates the best-fit to the data from \cite{DiazSantos2017} using Equation \ref{eq:lineFit}. Dotted lines correspond to $\pm1\sigma$ about the best-fit. (\textit{Right}) The ratio of \lcii\ to \lpah\ (summing over all features between $6-18\,\mu m$) for star-forming (U)LIRGs with \textit{Hershel}/PACS \cii\ measurements and \textit{Spitzer}/IRS observations of the PAHs. To compare our results with more normal star-forming conditions, we show measurements of the spatially resolved nuclear ring and extranuclear bar in NGC 1097 as open symbols (squares and circles respectively) from \cite{Beirao2012} on both panels, as well as the range in \cii/\lir\ and \cii/PAH observed in the normal star-forming galaxy sample of \cite{Helou2001} and \cite{Malhotra2001} (shaded grey regions), scaled by a factor of 2.3 in the \textit{Right} panel as described in Figure \ref{fig:cii_pah_vs_lir}. Both \cii/\lir\ and \cii/PAH trend towards lower values in more compact systems, and the use of PAH emission as a normalization for \cii\ yields a lower scatter about the overall trend. }
    \label{fig:deficitFig}
\end{figure*}


\begin{figure*}
    \centering
    \includegraphics[width=\textwidth]{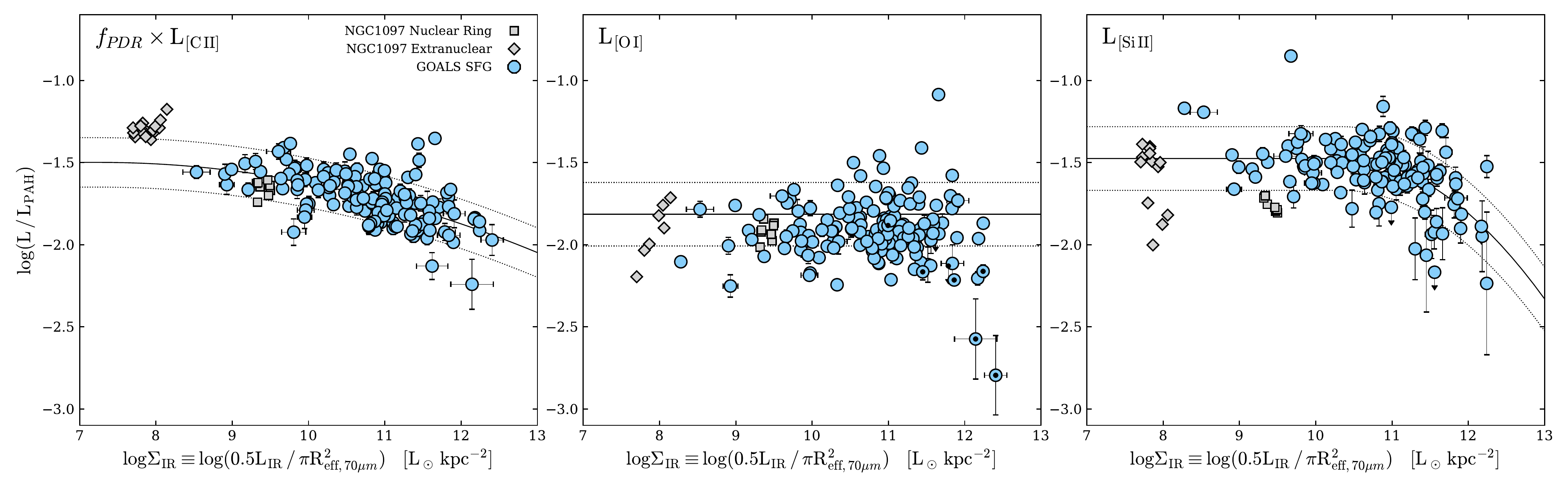}
    \caption{IR cooling lines over PAH emission in GOALS star-forming (U)LIRGs vs. IR surface densities. On all panels, we show the best-fit of Eq. \ref{eq:lineFit} to the data (solid black line), where all parameters are left free. Dotted black lines indicated $\pm1\sigma$ scatter about the trend. (\textit{Left}) \cii/PAH with the ionized component of \cii\ subtracted out using the \cii/[\ion{N}{2}]$_{205}$ and [\ion{N}{2}]$_{122}$/[\ion{N}{2}]$_{205}$ ratio as described in \cite{DiazSantos2017}. (\textit{Center}) \oi/PAH. Galaxies with signs of \oi\ self-absorption are marked with black dots, which comprise 5\% of the sample.  (\textit{Right}) \siii\ $34.8\,\mu m$ over PAH. In all panels, we compare GOALS star-forming galaxies to the  star-forming, compact nuclear ring, and extranuclear emission from NGC 1097 presented in \cite{Beirao2012}. The spatially resolved emission in a normal star-forming galaxy is consistent with extrapolation of the trends observed in LIRGs and ULIRGs to low \sigmaIR. While the fraction of cooling-to-heating in \cii\ falls in the most compact systems, \oi/PAH remains relatively constant over \sigmaIR\ in (U)LIRGs. The lack of a positive trend between \oi/PAH and \sigmaIR\ suggests that \oi\ cooling does not compensate for the negative trends observed in the other features. }
    \label{fig:line_over_pah}
\end{figure*}


\begin{figure*}
    \centering
    \includegraphics[width=.95\textwidth]{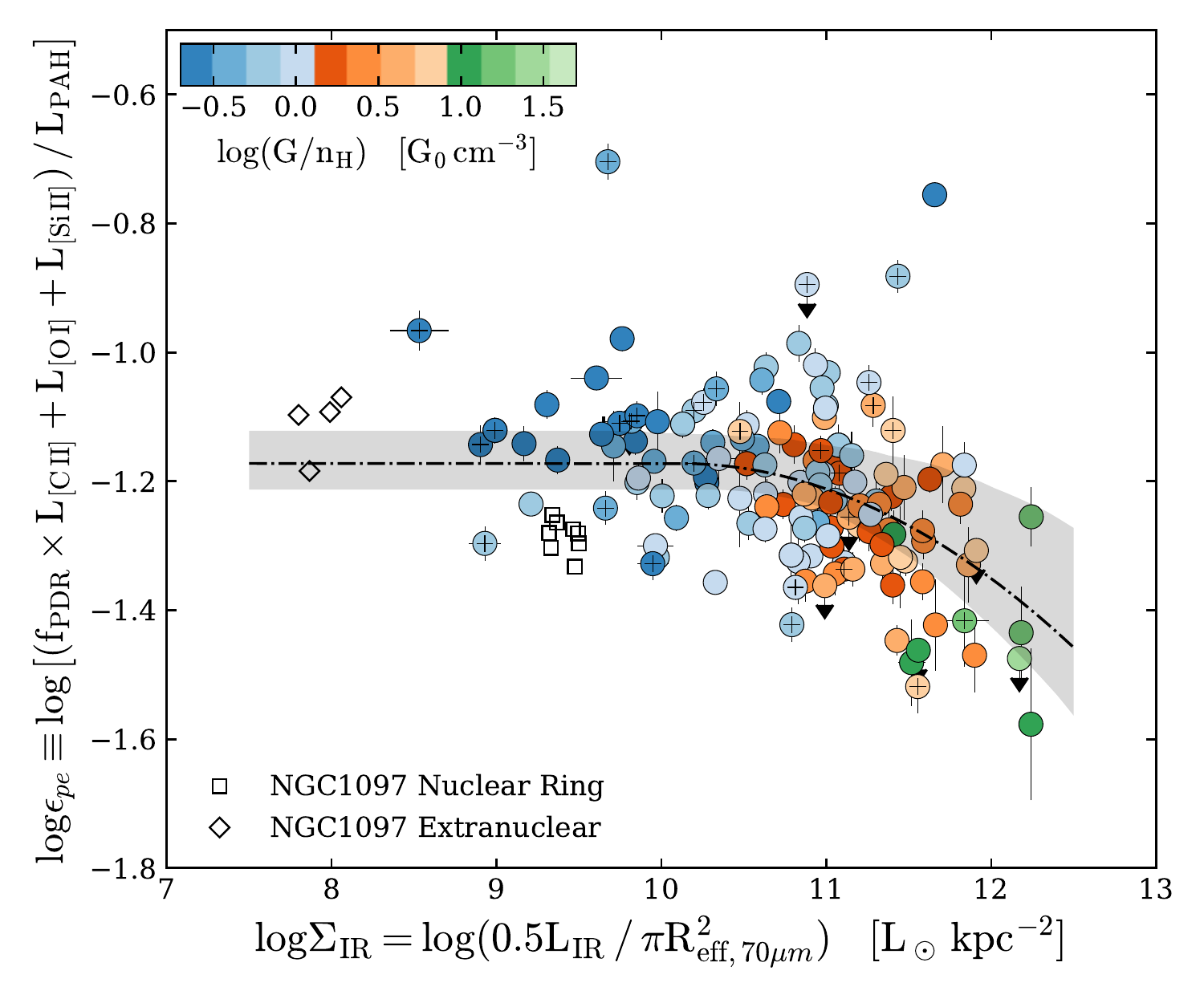}
    \caption{\small The ratio of prominent IR cooling lines to total PAH luminosity, an estimate of the photoelectric efficiency in PDRs, vs. IR surface density. All measurements along the $y$-axis have been aperture-matched as discussed in Section \ref{sec:DataAnalysis}. The data are color-coded by $\log(\mathrm{G/n_H)}$, following previous Figures. The dash-dotted line corresponds to the best-fit of Eq. \ref{eq:lineFit} to the data, with all parameters set to free. The shaded grey region contains the top 95\% of 1000 boot-strapped model fits.  Data points marked with a $+$ symbol have mid-IR AGN fractions $>30\%$, and low $6.2\,\mu m$ equivalent widths, either due to a deeply buried starburst or AGN. For comparison with more normal star-forming conditions, we show measurements of the spatially resolved nuclear ring and extranuclear bar in NGC 1097 as open symbols (squares and circles respectively) from \cite{Beirao2012}. The Kendall's $\tau$ and two-tailed $\log\,p-$value for the (U)LIRGs are $-0.3$ and $-7.1$ respectively, indicating an anti-correlation between \ephot\ and \sigmaIR\ at a confidence level of $\approx5\sigma$ (Tab. \ref{tab:corr}), driven largely by the turn-over in \ephot\ at $\log$\sigmaIR$\,>10.5$ in galaxies with $\log(\mathrm{G/n_H})\gtrsim0.5$. At these high surface densities, the average physical properties of PDRs depart from the more normal star-forming conditions observed at lower \sigmaIR. }
    \label{fig:ephot}
\end{figure*}

\section{Results\label{sec:results}}
Following early \cii\ surveys of low-redshift, mostly normal star-forming galaxies with \textit{ISO}, \cite{Helou2001} measured a constant \cii/PAH ratio over a range of far-IR colors that showed significantly less scatter than \cii/\lir. This reinforced the link between PAHs and \cii\ emission via photoelectric heating inside PDRs. We find that the galaxy-integrated \cii/PAH ratio in luminous, infrared, star-forming galaxies does not correlate strongly with either far-IR color or \lir\ as shown in Figure \ref{fig:cii_pah_vs_lir}, although warmer LIRGs and ULIRGs tend to have lower \cii/PAH ratios compared to galaxies at \loglir$\lesssim11$. Trends in PDR gas cooling/heating within the GOALS sample are likely to be more clear when measured against quantities that better reflect the ionizing radiation or luminosity density surrounding the young stars.

\begin{deluxetable}{lcccc}
    \tablecaption{Best-fit Parameters to to Line/PAH ratios vs. \sigmaIR \label{tab:par}}
    \tablehead{Line & $a_0$ & $a_1$ &  $a_2$ & $1\sigma$ (dex) }
    \startdata 
    \cii & -0.278 & -0.124 & 0.00 & 0.08 \\[.3ex]
    \fpdr$\times$\cii & -2.378 & 0.242 & -0.016 & 0.15 \\[.3ex]
    \oi\ & -1.997 & 0.013 & 0.00 & 0.19 \\[.3ex]
    \siii\ & -14.325 & 2.487 & -0.120 & 0.19  \\[.3ex]
    \ephot\tablenotemark{a}\ & -6.277 & 1.010 & -0.050 & 0.12  \\[.3ex]
    \enddata
    \tablecomments{All line luminosities have been calculated from aperture matched flux measurements as described in Section \ref{sec:DataAnalysis} unless otherwise noted. Fits are a function of $\log_{10}$\sigmaIR$\,=\log_{10}\,$(0.5\lir$/\pi R_{eff,70\mu m}^2)$ in units of L$_\odot$ kpc$^{-2}$. The right-most column corresponds to the $1\sigma$ scatter about the trend. }
    \tablenotetext{a}{See Equation \ref{eq:ephot}.}
  \end{deluxetable}

\subsection{Line-to-PAH Ratios}
The IR surface density is a more direct tracer of far-IR fine-structure line properties than color or \lir\ \citep{DiazSantos2013,DiazSantos2017,Smith2017}, both of which exhibit little to no influence on the ratio of \cii/PAH (Fig. \ref{fig:cii_pah_vs_lir}). As \sigmaIR\ should scale with the number density of massive star-forming regions in the beam, we expect \sigmaIR\ to be more sensitive to statistical trends across the sample. The \cii\ deficit vs. \sigmaIR\ among luminous infrared star-forming galaxies presented in \cite{DiazSantos2017} is re-created for comparison in Figure \ref{fig:deficitFig} (\textit{Left}). To test for trends in gas cooling over photoelectric heating, we calculate the \cii/PAH ratio as a function of \sigmaIR, shown in Figure \ref{fig:deficitFig} (\textit{Right}), color coded by $\log\,\mathrm{G/n_H}$ which scales linearly with IR surface density above \logSigmaIR$\sim10.7$ \citep{DiazSantos2017}. Whereas the \cii/\lir\ ratio exhibits a maximum deficit of $\sim2$ dex between \logSigmaIR$\,=8-13$, the \cii/PAH ratio falls by a factor of $\sim0.5$ dex over the same range. LIRGs at low \sigmaIR\ exhibit \cii/\lir\ and \cii/PAH ratios comparable to those found in the spatially-resolved regions of NGC 1097 \citep{Beirao2012}, and the normal star-forming galaxies in \cite{Helou2001}.

Following \cite{DiazSantos2017}, we fit a second-order polynomial function to the \cii/PAH ratio of the form: 
\begin{equation}
    \log(\mathrm{L_{line}/L_{IR}})=a_0 + a_1\,\log\Sigma_{\mathrm{IR}}+a_2\,(\log\Sigma_{\mathrm{IR}})^2
    \label{eq:lineFit}
\end{equation}
Note that the function is forced to a constant maximum at values of \sigmaIR\ less than where the turn-over occurs. While we do not include normal star-forming galaxies in the fit, a constant line-to-PAH ratio below the turn-over yields the best agreement with observations of low \sigmaIR\ galaxies (e.g., \citealt{Helou2001,Beirao2012,Croxall2012}). The best-fit is shown as a black solid line in Fig. \ref{fig:deficitFig} (\textit{Right}), the parameters for which are given in Table \ref{tab:par}. While the \cii/PAH ratio is not constant, the magnitude of the drop with \sigmaIR\ is significantly less. 

In addition to \cii, \oi\ and \siii\ are important PDR cooling lines in terms of their overall contribution to the cooling budget (e.g., \citealt{Rosenberg2015}), and their strengths relative to the PAH emission are a diagnostic of PDR structure. Large \oi/PAH ratios where \cii/PAH is low could indicate greater gas densities, which would preferentially cool through the higher critical density \oi\ line. In Figure \ref{fig:line_over_pah}, we show the \cii/PAH, \oi/PAH, and \siii/PAH ratios in separate panels for LIRGs and ULIRGs as a function of \sigmaIR. Because we are comparing different fine-structure cooling lines here and would like to focus on the PDR emission alone, we  subtract out the non-PDR component of the \cii\ emission using \fpdr\ (e.g., \citealt{DiazSantos2017}). At low IR surface densities, the line-to-PAH ratios are consistent with spatially-resolved measurements of NGC 1097 \citep{Beirao2012}, representative of more normal star-forming conditions. 

\cii/PAH falls in the most compact sources at \logSigmaIR$\,\gtrsim10$. \oi/PAH does not exhibit a turn-over at high \sigmaIR, but shows an increasingly larger scatter (higher than the other line ratios), mostly with a constant lower envelope. This could arise from varying degrees of line self-absorption, as \oi\ originates from deeper regions within PDRs, and/or the presence of intervening (foreground) cold, neutral material. Indeed, IRASF1224-0624, and ICO860, the two galaxies $>2\sigma$ below the \oi/PAH trend both show signs of self-absorption in the PACS spectra and have some of the highest $\tau_{9.7}$ optical depths observed in the sample; however, \oi\ self-absorption is only observed in $\sim5\%$ of GOALS in total (12 objects, see \citealt{DiazSantos2017}) spanning \logSigmaIR$=10-12.5$, and is therefore unlikely to influence statistical trends in the sample over the full range in \sigmaIR.

\begin{deluxetable}{lrcc}
    \tablecaption{Correlation Coefficients between the Photoelectric Efficiency, as defined by the ratio of IR cooling lines to mid-IR PAH emission between $6-12\,\mu m$, and other Quantities in GOALS star-forming Galaxies \label{tab:corr}}
    \tablehead{ & $\tau_k$ &  $\log p$ & SNR [$\sigma$\tablenotemark{a}] }
    \startdata 
    \lir$\,\mathrm{[L_\odot]}$ & $-0.15_{-0.2}^{-0.1}$ & $-2.21_{-3.7}^{-1.1}$ & $2.7\,^{3.7}_{1.7}$ \\[.7ex]
    $S_{\nu,\,63\mu m}/S_{\nu,\,158\mu m}$ & $0.00_{-0.09}^{0.08}$ & $-0.35_{-0.91}^{-0.09}$ & $0.8\,^{1.5}_{0.2}$ \\[.7ex]
    \sigmaIR$\,\mathrm{[L_\odot\,kpc^{-2}]}$\tablenotemark{b} & $-0.30_{-0.4}^{-0.2}$ & $-7.1_{-9.5}^{-4.9}$ & $5.3\,^{6.3}_{4.4}$ \\[.7ex]
    $\mathrm{G/n_H}\,\mathrm{[G_0\,cm^{-3}]}$\tablenotemark{c} & $-0.31_{-0.35}^{-0.26}$ & $-9.0_{-11.6}^{-6.6}$ & $6.1\,^{7.0}_{5.2}$ \\[.7ex]
    EW($6.2\,\mu m$)$\,\mathrm{[\mu m]}$ & $0.08_{0.01}^{0.1}$ & $-0.8_{-1.9}^{-0.2}$ & $1.5\,^{2.5}_{0.4}$ \\[.7ex]
    \fagn & $0.08_{0.03}^{0.1}$ & $-0.9_{-2.1}^{-0.3}$ & $1.6\,^{2.6}_{0.6}$ \\[.7ex]
    \fpdr & $0.06_{0.0}^{0.1}$ & $-0.6_{-1.5}^{-0.1}$ & $1.1\,^{2.2}_{0.3}$ \\[.7ex]
    $\mathrm{L_{11.3\,\mu m}/L_{7.7\,\mu m}}$ & $0.06_{0.0}^{0.12}$ & $-0.6_{-1.7}^{-0.2}$ & $1.2\,^{2.3}_{0.4}$ \\[.7ex]
    $\mathrm{L_{7.7\,\mu m}/L_{6.2\,\mu m}}$ & $-0.23_{-0.3}^{-0.2}$ & $-4.9_{-7.1}^{-3.1}$ & $4.3\,^{5.4}_{3.3}$ \\[.7ex]
    $\mathrm{L_{11.3\,\mu m}/L_{3.3\,\mu m}}$\tablenotemark{d} & $-0.37_{-0.42}^{-0.32}$ & $-7.5_{-9.5}^{-5.7}$ & $5.5\,^{6.2}_{4.8}$ \\[.7ex]
    \enddata
    \tablecomments{Kendall's $\tau$ correlation coefficients ($\tau_k$) and $p-$values were calculated using \texttt{pymccorrelation} \citep{Privon2020}, which implements bootstrap error estimation on $\tau_k$ and $p$ with censored data (upper limits) as described in \cite{Curran2014} and \cite{Isobe1986} respectively. We report 16\%, 50\%, and 84\% percentiles for each quantity. }
    \tablenotetext{a}{ Assuming normally distributed posteriors.}
    \tablenotetext{b}{ See Figure \ref{fig:ephot}.}
    \tablenotetext{c}{ See Figure \ref{fig:ephot_gn}.}
    \tablenotetext{d}{ For the subset of GOALS (U)LIRGs with AKARI 3.3$\,\mu m$ PAH detections. See Figure \ref{fig:ephot_vs_1133}.}
  \end{deluxetable}


\subsection{The Photoelectric Efficiency\label{sec:results_ephot}}
The photoelectric efficiency (\ephot) is the fraction of energy that photoelectrons from PAHs transfer from UV photons into the gas relative to the total heating of dust.~Observationally, \ephot\ can be traced by the ratio of far-IR line emission to total PAH emission because photoelectrons from PAHs are one of the dominant heating mechanisms in PDRs, which cool predominantly via far-IR fine-structure line emission (e.g.,   \citealt{Bakes1994,Malhotra2001,Croxall2012,Beirao2012}). In practice, the cooling budget is dominated by \cii, \oi, and \siii, and \lpah$\gg$(\lcii+\loi+\lsiii).~Assuming that the PAHs and ions are co-spatial such that the energy input into the gas by photoelectric heating powers the \cii, \oi, and \siii\ lines, an empirical tracer of the photoelectric efficiency is  
\begin{eqnarray}
    \epsilon_{\mathrm{PE}}\approx\frac{\mathrm{L_{[C\,II]}+L_{[O\,I]}+L_{[Si\,II]}}}{\mathrm{L_{PAH}}}
    \label{eq:ephot}
\end{eqnarray}
where the relative contribution of far-IR lines to \ephot\ can vary across the PDR \citep{Kaufman2006}. Nevertheless, combined observations of these features capture an average \ephot\ of PDRs over a galaxy. 

We measure the average \ephot\ for 152 GOALS star-forming galaxies including five upper limits, and show the combined sum of neutral \cii, \oi, and \siii\ cooling over \lpah\ in Figure \ref{fig:ephot}.  A Kendall's $\tau$ test supports an anti-correlation between \ephot\ and \sigmaIR\ at $\sim5\sigma$ (see Table \ref{tab:corr}). 
As shown in Fig.~\ref{fig:ephot}, LIRGs at low \sigmaIR\ exhibit fairly constant photoelectric efficiencies comparable to observations of normal star-forming galaxies \citep{Beirao2012,Croxall2012}. At high \sigmaIR, the sum of all cooling lines relative to the PAHs shows a decreasing photoelectric efficiency, which suggests that the net cooling to photoelectric heating ratio is falling across the density and temperature structure of PDRs in compact objects. The relative contributions of \cii, \oi, and \siii\ to the cooling budget within PDRs (i.e., the numerator of Eq. \ref{eq:ephot}) are on average 32\%, 20\%, and 48\% respectively, and these values do not change significantly when splitting the sample above and below \logSigmaIR$=10.7$. Therefore, low photoelectric efficiencies are the product of overall less cooling in PDRs in all channels relative to PAH heating. 

The anti-correlation between \ephot\ and \sigmaIR\ is robust against outliers, as demonstrated by the bootstrap fits on Figure \ref{fig:ephot}. Nevertheless, there exist galaxies deviating from the trend by more than $\pm2\sigma$ that warrant closer inspection. At all IR surface densities, objects in this class collectively have higher mean and median \fagn\ compared to the rest of the sample. Thus, the scatter above and below the best-fit trend may be partially due to sources with an excess of hot dust continuum emission in their mid-IR spectra. This may be due to a weak, deeply buried AGN, even in sources where \lir\ is dominated by star formation, which in turn may change the ionization structure and the PAH properties \citep{Voit1992,Langer2015,Langer2017}. Alternatively, these objects may harbor heavily obscured star-forming regions with high levels of line self-absorption and/or continuum opacities. This absorption should appear in the PACS line profiles, but only 12 GOALS sources show evidence for absorption in the \oi\ line \citep{DiazSantos2017}. We note that the detection of these signatures, presumably due to compact (narrow velocity) foreground clouds, may be limited by the coarse spectral resolution ($\sim85\,\mathrm{km\,s^{-1}}$) of \textit{Herschel}/PACS \citep{Gerin2015}. 

\begin{figure}[h!]
    \centering
    \includegraphics[width=.48\textwidth]{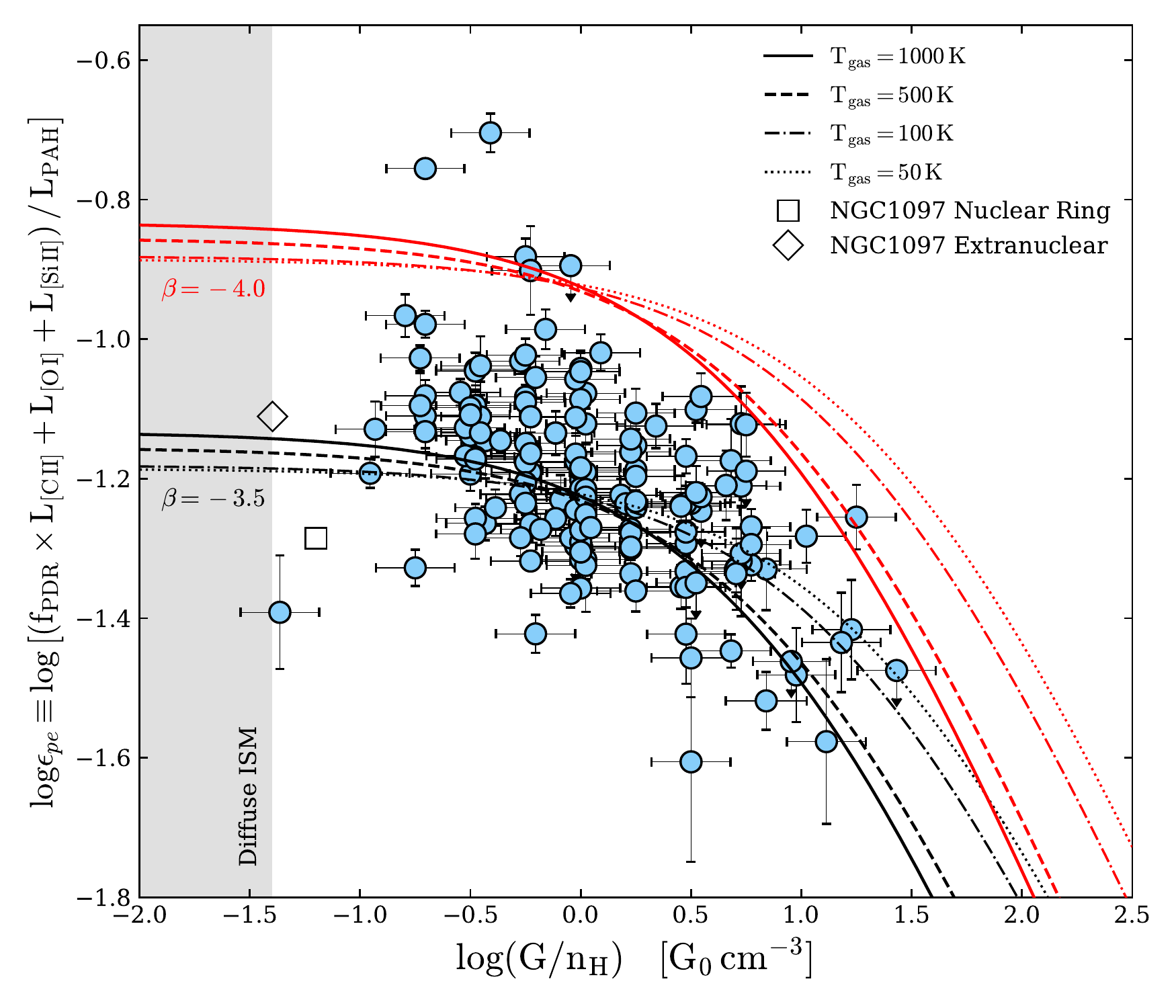}
    \caption{The photoelectric efficiency calculated as the sum of \cii, \oi, and \siii\ emission over the flux of PAH emission including lines between $6-18\,\mu m$ vs. $\mathrm{G/n_H}$, the ratio of the average radiation field strength impinging upon PDRs to the mean neutral Hydrogen density. The curves correspond to theoretical photoelectric efficiencies of PAH grains from \cite{Bakes1994} for gas temperatures between $50-1000$ K, and for grain size distributions of the form $n(a)\sim a^{\beta}$. Black and red curves indicate $\beta=-3.5$ and $\beta=-4.0$ respectively, with the more negative exponent corresponding to a distribution weighted more towards smaller grains. Note that the \lpah\ we use to measure \ephot\ includes only $\sim75\%$ of the total power in PAHs for GOALS (see Sect. \ref{sec:pahprops}), and we have scaled the models accordingly to account for this difference. The shaded grey region corresponds to $\mathrm{G/n_H}$ typical of the cold, diffuse ISM. Open symbols correspond to the average values of \ephot\ and $\mathrm{G/n_H}$ for spatially resolved measurements of the nuclear ring and extranuclear bar (square and circle respectively) in NGC 1097, a more normal star-forming galaxy \citep{Beirao2012}. The models, which depend on the PAH grain ionization state and size distribution, can account for both the overall trend and scatter observed in (U)LIRGs, suggesting that the strong correlation between the photoelectric efficiency and $\mathrm{G/n_H}$ originates from the photoelectric properties of the PAH grains within PDRs.}
    \label{fig:ephot_gn}
\end{figure}

The photoelectric efficiencies we measure represent averages over the local properties of PDRs in the central ($\sim2-7$ kpc) star-forming regions of luminous infrared galaxies, and the trend in Fig. \ref{fig:ephot} suggests a departure from the typical ISM and PDR conditions at high \sigmaIR. Indeed, galaxies in this domain with low \ephot\ also have the largest ratios of G/n$_\mathrm{H}$, which is also sensitive to the local physics of PDRs. We fit Equation \ref{eq:lineFit} to the data, the parameters of which are given in Table \ref{tab:par}, and find that the photoelectric efficiency turns over at \logSigmaIR$\sim10.5$, close to where G/n$_\mathrm{H}$ scales linearly with $\log$\sigmaIR\ \citep{DiazSantos2017}. Models of heating and cooling in PDRs predict an anti-correlation between G/n$_\mathrm{H}$ and \ephot\ as the size and charge of dust grains are modified by the stellar radiation field (e.g., \citealt{Bakes1994,Tielens2008}), a trend which we recover in Figure \ref{fig:ephot_gn} at high ($\sim6\sigma$) significance (see Tab. \ref{tab:corr}). We find general agreement between modeled and measured \ephot\ for $\mathrm{T_{gas}}=50-1000$ K, reasonable values for temperatures across the phase structure of PDRs (e.g., \citealt{Tielens1985,Bakes1994,Draine1996}), and the two sets of models are able to account for the range in \ephot\ we measure in GOALS. The most extreme ULIRGs at high \sigmaIR\ and low \ephot\ also show the highest values of $\log\,(\mathrm{G/n_H})\gtrsim0.25$ supporting a link between the strength of the radiation field and the heating efficiency within the PDR, which is mediated by the properties of PAH grains. 


\begin{figure*}
    \centering
    \includegraphics[width=\textwidth]{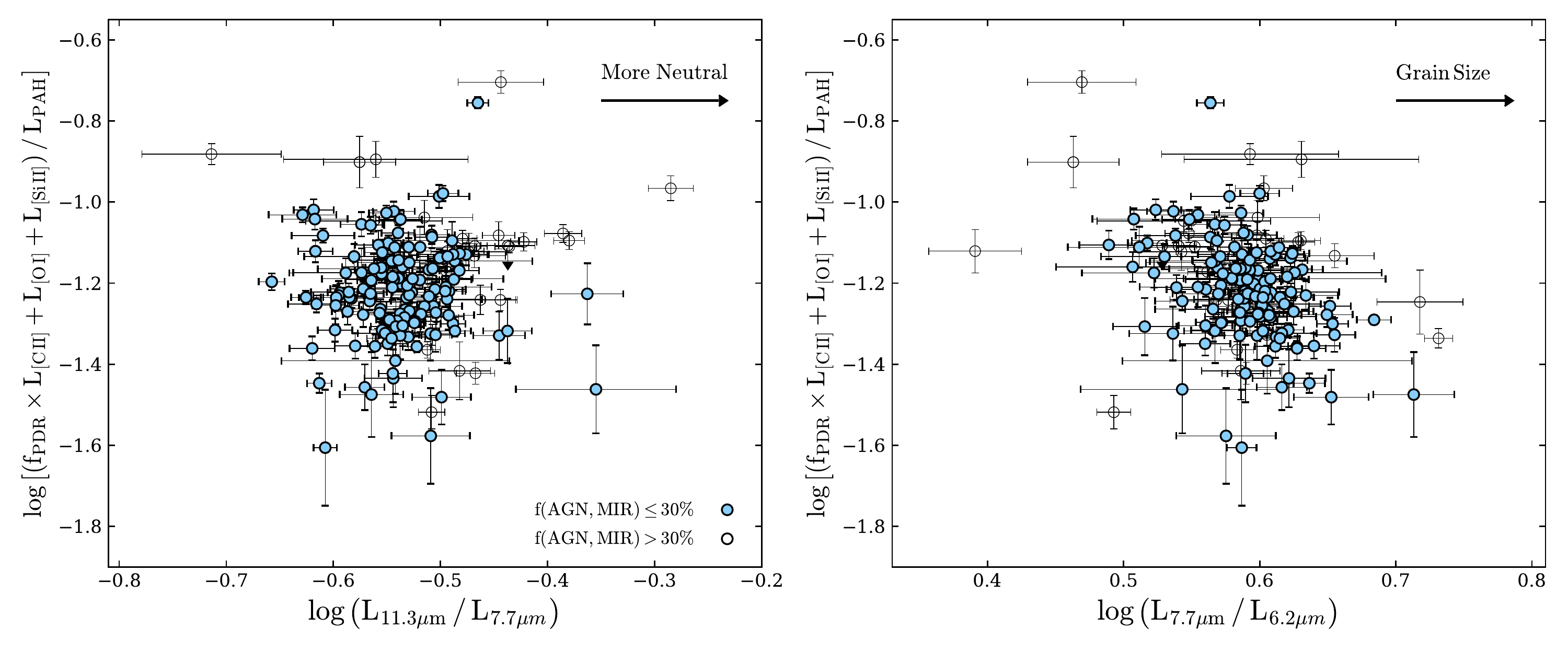}
    \caption{(\textit{Left}) The photoelectric efficiency vs. the ratio of 11.3 to 7.7$\,\mu m$ PAH emission in our sample of star-forming GOALS (U)LIRGs. The \LelevenSeven\ PAH ratio is a tracer of the average grain ionization, with higher values associated with more neutral grains. (\textit{Right}) Same as the \textit{Left} panel, now as a function of the ratio of 7.7 to 6.2$\,\mu m$ PAH emission, a tracer of the average Cationic grain size as both features arise from ionized PAHs. The low values of \ephot\ observed at high \sigmaIR\ and high $\mathrm{G/n_H}$ do not exhibit notably different grain charge states and Cation grain sizes. }
    \label{fig:ephot_vs_lineRatios}
\end{figure*}


\section{Discussion\label{sec:discussion}}

In this work, we find that the cooling-to-heating ratios observed in the most compact luminous infrared galaxies are low, and are accompanied by an increase in the  mean energy density per H atom ($\mathrm{G/n_H}$) incident upon the surfaces of PDRs. The IR surface density is a global property of a galaxy, and both \ephot\ and $\mathrm{G/n_H}$ measure the average physical conditions local to the PDRs. The trends between \ephot, $\mathrm{G/n_H}$, and \sigmaIR\ suggest a link between the global properties of a starburst and the conditions of individual star-forming regions. 
\newline 

\subsection{Alternate Explanations for decreasing \ephot\ with \sigmaIR}
The low photoelectric efficiencies we measure in galaxies with high \sigmaIR\ (Fig.~\ref{fig:ephot}) and high $\mathrm{G/n_H}$ (Fig.~\ref{fig:ephot_gn}) suggests a change in the thermal regulation of star-forming gas in compact LIRGs; however, a number of alternative physical conditions may conspire to produce the trends we observe with \ephot.

The relative strengths of the far-IR cooling or PAH emission may be affected by any or all of the following: (1) Far-IR line fluxes may be suppressed by optically thick continuum absorption (e.g., \citealt{Scoville2017b}). In this case, the \oi/\lir-deficit would have a steeper slope than \cii/\lir\ (e.g., \citealt{Malhotra2001}), a trend that is not observed \citep{DiazSantos2017}. (2) Ionized gas from diffuse or \ion{H}{2} regions, and/or AGN may contribute to the \cii\ and \siii\ line fluxes owing to their ionization potentials being lower than that of neutral hydrogen. This is unlikely to drive trends in our sample, as the photoelectric efficiency does not correlate with the fraction of \cii\ emission from PDRs (\fpdr; Tab. \ref{tab:corr}). Moreover, the neutral PDR fraction of \siii\ would have to be larger at low \sigmaIR\ or smaller at high \sigmaIR\ to maintain constant \ephot, both of which are inconsistent with the increase in PDR area filling factor with \sigmaIR\ in LIRGs and ULIRGs \citep{DiazSantos2017}. (3) PAH molecules in galaxies with large $\mathrm{G/n_H}$ absorb and re-emit more energy per unit dust mass because of the large UV opacities of grains \citep{Li2001}. If the PAHs and ions (C$^+$, O, and Si$^+$) were de-coupled spatially, \lpah\ could increase with \sigmaIR\ without a corresponding increase in (\lcii+\loi+\lsiii), lowering the observed IR line-to-PAH ratio by Eq.~\ref{eq:ephot}. However, this spatial decoupling of PAHs and ions is inconsistent with observations of PDRs in the Milky Way \citep{Okada2013,Salgado2016}, the LMC \citep{Lebouteiller2012,Chevance2016}, and local galaxies \citep{Croxall2012,Abdullah2017,Bigiel2020}.~(4) Metallicity may influence the strength of cooling lines and the PAHs \citep{Cormier2015,Smith2017,Croxall2017,Cormier2019,Aniano2020}. However, direct effects on the PAH grains appear most pronounced at metallicities well below those seen in GOALS galaxies, which typically have $\mathrm{Z/Z_\odot\sim1-2}$ \citep{Inami2013,DiazSantos2017}. Therefore, the trends observed in Fig.~\ref{fig:ephot} and Fig.~\ref{fig:ephot_gn} likely reflect variations in the heating and cooling mechanisms on the scale of individual PDRs.

\subsection{Dust Grain Properties and the Photoelectric Efficiency}
The photoelectric efficiency is not only a function of $\mathrm{G/n_H}$, but also the size and ionization state of PAH grains \citep{Bakes1994,Galliano2008,Tielens2008}. Observations of PDRs in the Milky Way indicate that \ephot\ falls as the grain charging parameter ($\gamma\equiv\mathrm{G_0 T^{1/2}/n_e}$) rises, and grains become more ionized \citep{Tielens2008,Okada2013,Salgado2016}. Indeed, the models of \cite{Bakes1994} shown in Figure \ref{fig:ephot_gn} predict low \ephot\ at high $\mathrm{G/n_H}$, and the data falls closer to the model with a grain size distribution weighted more towards larger grains. Thus, a link between PAH properties and \ephot\ is to be expected if the decrease in \ephot\ at high \sigmaIR\ arises from a change in the local physics of gas heating in PDRs as mediated by the charge and size of PAHs.

Ratios between the 6.2, 7.7, 8.6, and 11.3$\,\mu m$ PAH lines are well-established as tracers of grain size and ionization (e.g., \citealt{Draine2001,Tielens2008}), and local LIRGs exhibit comparable ratios to nearby, normal star-forming galaxies and high-$z$ ULIRGs \citep{Stierwalt2014,Smith2007,Pope2008,Wu2010,Shipley2013}. For example, star-forming LIRGs cluster tightly in \LsixSeven\ vs. \LelevenSeven, tracing grain size and ionization respectively, which show a larger scatter in sources with AGN \citep{Stierwalt2014}.  

Figure \ref{fig:ephot_vs_lineRatios} shows \ephot\ as a function of the \LelevenSeven\ ratio, a tracer of grain charge, and the \LsevenSix\ ratio, a tracer of the average cationic grain size \citep{Draine2001}. We do not detect a correlation between \ephot\ and the grain ionization state at a significant confidence level (Tab. \ref{tab:corr}); however, \ephot\ and \LsevenSix\ are anti-correlated at a $\sim4\sigma$ confidence level, albeit with large scatter, suggesting that that the average size of PAH grains may influence the observed photoelectric efficiencies in local (U)LIRGs, consistent with the sensitivity of theoretical photoelectric efficiencies on the grain size distribution (Fig. \ref{fig:ephot_gn}).  

Smaller PAH grains emit more strongly in the shorter wavelength bands, whereas larger grains are brighter at longer wavelengths \citep{Allamandola1989,Schutte1993,Draine2001}. Thus, the diagnostic utility of a PAH line ratio as a grain-size indicator is a function of the separation in wavelength of the two features. The 7.7/6.2$\,\mu m$ ratio has been a wide-spread tool in the literature as both features are bright in extragalactic sources, relatively unaffected by the 9.7$\,\mu m$ silicate ice feature, and were observable simultaneously with IRS aboard \textit{Spitzer} \citep{Draine2001,ODowd2009,Sandstrom2012,Stierwalt2014}. With the advent of AKARI/IRC, the 3.3$\,\mu m$ PAH feature became accessible for large numbers of extragalactic sources while unlocking a longer baseline diagnostic of grain size. 

\begin{figure}
    \centering
    \includegraphics[width=.47\textwidth]{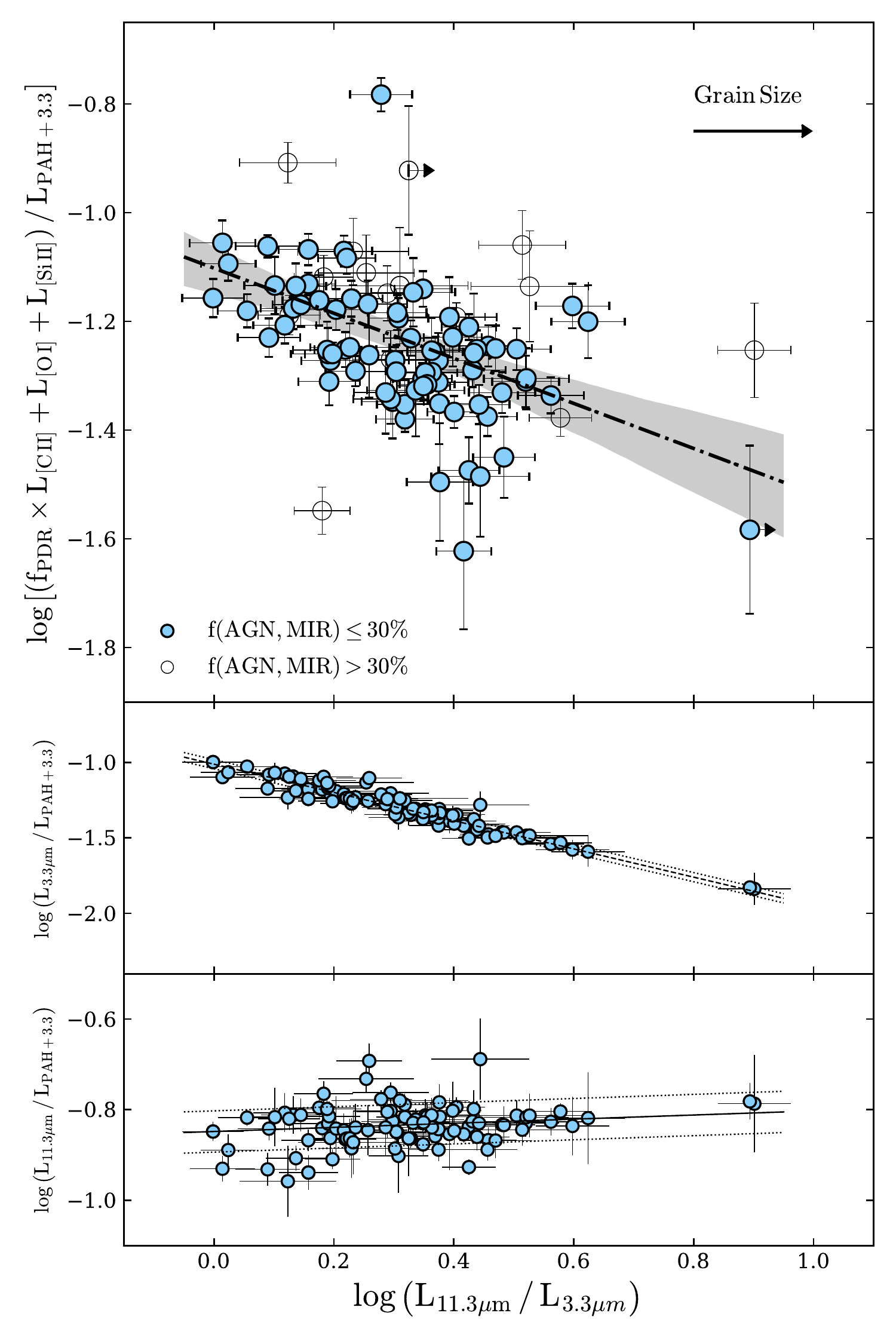}
    \caption{(\textit{Top}) The photoelectric efficiency vs. the ratio of $11.3\,\mu m$ to $3.3\,\mu m$ PAH emission, a size tracer for PAH grains. Note that $\mathrm{L_{3.3}}$ is not included in \lpah. (U)LIRGs with higher \ephot\ are low in 11.3/3.3$\,\mu m$. We fit a linear relation to the data letting the slope and $y-$intercept vary freely, and find that $y=-0.41x-1.10$. The shaded grey region spans the domain of 1000 bootstrapped model fits. A Kendall's $\tau$ test returns $(\tau_k,\log\,p)=(-0.37,-7.5)$, indicating an anti-correlation at $\sim5.5\sigma$ confidence. Galaxies with \fagn$>30\%$ (open circles) exhibit a $1\sigma$ scatter about the best-fit of $0.11$ dex, larger than the $0.07$ dex scatter in low \fagn\ (U)LIRGs ($\leq30\%$, filled circles).  (\textit{Center}) The $\mathrm{L_{3.3}/L_{PAH}}$ ratio as a function of the 11.3/3.3$\,\mu m$ ratio. (\textit{Bottom}) The contribution of the $11.3\,\mu m$ PAH feature to the total PAH luminosity. (U)LIRGs with lower photoelectric efficiencies tend towards higher values of 11.3/3.3 PAH ratios without exhibiting changes to the fractional strength of the $11.3\,\mu m$ feature with respect to \lpah, supporting a link between \ephot\ and the PAH grain-size distribution probed by the 3.3$\,\mu m$ feature luminosity.  }
    \label{fig:ephot_vs_1133}
\end{figure}

The ratio of $11.3\,\mu m$ to $3.3\,\mu m$ PAH intensity is one of the most robust tracers of PAH grain size because of the large baseline in wavelength \citep{Allamandola1989,Jourdain1990,Schutte1993,Mori2012,Ricca2012,Croiset2016,Maragkoudakis2020}. Using the available AKARI spectra in GOALS, we plot \ephot\ vs. \LelevenThree\ in Figure \ref{fig:ephot_vs_1133} to further test if low \ephot\ is associated with larger or smaller PAHs. We find that galaxies in the AKARI sub-sample of GOALS with low \ephot\ show systematically higher values of 11.3/3.3$\,\mu m$ PAH emission (Tab.~\ref{tab:corr}). This trend is not driven by systematic variations in the 11.3$\,\mu m$ feature strength, as the $\mathrm{L_{11.3\,\mu m}}$/\lpah\ ratio is constant over the range in \LelevenThree, and exhibits minimal $1\sigma$ scatter of 0.05 dex about the average (Fig.~\ref{fig:ephot_vs_1133}, bottom panel). Instead, the strength of the $3.3\,\mu m$ features relative to \lpah\ decreases dramatically by nearly an order of magnitude (Fig.~\ref{fig:ephot_vs_1133}, center panel). The anti-correlation between \ephot\ and \LelevenThree\ reinforces the importance of the mean grain size in dictating the observed photoelectric efficiency. 


\subsection{Physical Interpretation}

The photoelectric heating efficiency of PAHs falls as the number of carbon atoms per particle increases (e.g., \citealt{Bakes1994}), and small grains are preferentially destroyed in the presence of harsh radiation fields (e.g., \citealt{Jochims1994}). This provides a simple physical interpretation of the data presented so far: PDRs in the most compact starbursts (\logSigmaIR$\,\gtrsim10.7$) have, on average, more intense radiation fields per gas density \citep{DiazSantos2017}, which leads to a preferential destruction of the smallest PAH grains, raising the average grain size and leaving behind large PAHs capable of releasing both less and weaker photo-electrons. This decreases the energy of a typical photoelectron available to heat the gas, and results in a lower gas heating efficiency.  

This scenario is consistent with principles of PDR evolution. The strength of the radiation field impinging upon PDRs (i.e., $\mathrm{G}$) is inversely proportional to the distance between the ionization front and the ionizing photon source \citep{Kaufman1999}. Therefore, higher values of $\mathrm{G/n_H}$ are associated with younger PDR systems that have yet to evolve away from their host star, and are more likely to be found in galaxies with higher star-formation rate surface densities (e.g., \citealt{DiazSantos2017}). The low photoelectric efficiencies at high \sigmaIR\ are plausibly linked to higher fractions of young, short-lived, and dust-cocooned star-forming regions, where the high ratios of $\mathrm{G/n_H}$ destroy small dust grains, hindering the coupling between the radiation field strength and PDR gas temperatures. Such PDRs may be continuously replenished by the compaction of gas and dust during a major merger \citep{Sanders1988,Hopkins2008}. Indeed, the majority of late-stage mergers (80\%) are above \logSigmaIR$\sim10.7$ \citep{Stierwalt2013,DiazSantos2017}, where photoelectric efficiencies are low. 

\begin{figure}
    \centering
    \includegraphics[width=.48\textwidth]{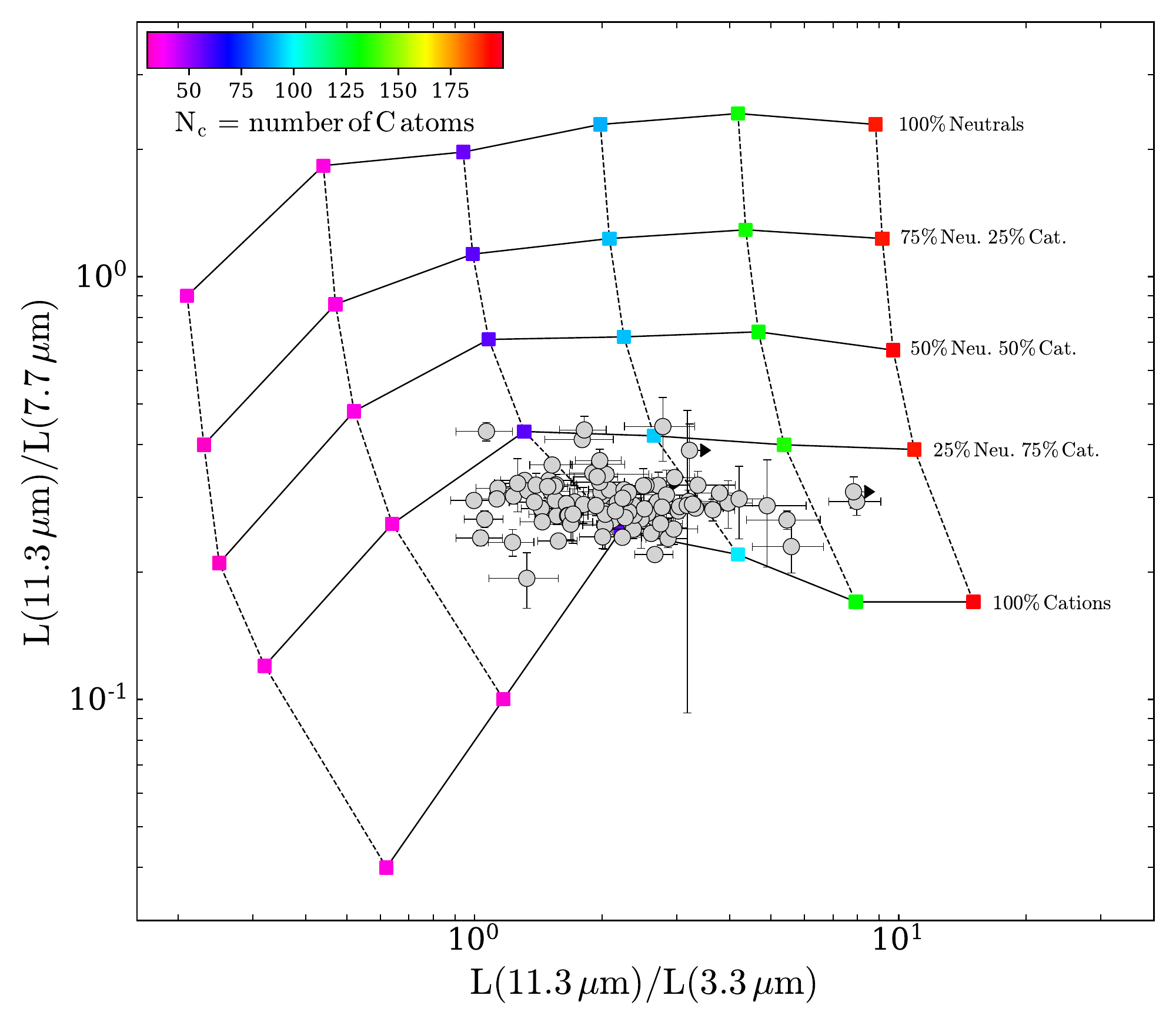}
    \caption{The \LelevenSeven\ vs. \LelevenThree\ PAH ratio in GOALS (U)LIRGs with AKARI observations of the 3.3$\,\mu m$ PAH feature. The model grid from \cite{Maragkoudakis2020} is a function of grain size ($\mathrm{N_C}$, see colorbar), and the relative mix of neutral and cationic PAH grains as indicated on the Figure, all assuming an average photon energy of 10 eV. (U)LIRGs have, on average, mostly ionized PAHs with sizes between 60 and 150 C atoms per grain.  }
    \label{fig:model}
\end{figure}

As shown in Figure \ref{fig:model}, the \LelevenSeven\ and \LelevenThree\ ratios in LIRGs and ULIRGs are consistent with predominantly ionized PAHs containing on average $\sim60-150$ C atoms per particle when compared to the modeled PAH spectra of \cite{Maragkoudakis2020}\footnote{We adopt the \cite{Maragkoudakis2020} models corresponding to a mean photon energy of 10 eV.}. PAHs with $\mathrm{N_C}\lesssim40$ tend to be photo-destroyed in PDRs \citep{Jochims1994}, which is close to the lower bound on grain sizes we observe for LIRGs and ULIRGs in Fig. \ref{fig:model}. Other mechanisms for grain destruction include shock-induced fragmentation and/or shattering; however, interstellar shocks tend to destroy larger grains which is the opposite effect we see in the data \citep{Jones1996}. Therefore, $\mathrm{G/n_H}$ may be a critical factor for determining the photoelectric efficiency in (U)LIRGs with the grain size distribution acting as an intermediary between the radiation field, and gas cooling and heating. 

\subsection{Prospects for Star-Formation at High-Redshift}

The rise and fall of the cosmic star-formation rate density may be accompanied by an increase in the efficiency of star-formation at earlier times, although this remains a subject of debate (e.g., \citealt{Lilly1996,Madau1996,MadauDickinson2014,Tacconi2010,Tacconi2013,Tacconi2018,Genzel2015,Scoville2017,Liu2019,Decarli2019}). Nevertheless, starbursts above the main-sequence have higher star-formation efficiencies (star-formation rate per unit molecular gas mass) than normal galaxies, independent of redshift \citep{Saintonge2013,Genzel2015,Scoville2017}. Can changes in the global star-formation efficiency of a galaxy, with redshift and/or distance from the main-sequence, be linked to variations in the ISM conditions on small scales? 

Given that photoelectric heating is the dominant coupling mechanism between gas temperatures and stellar radiation fields, \ephot\ could play a role in mediating the star-formation efficiency today and in the past by raising or lowering the energy transfer from stellar photons into the gas. When the photoelectric efficiency is low, photo-electrons convert a lower fraction of the incident radiation field into raising the gas temperature. In other words, gas may remain cold despite overall less cooling relative to photoelectric heating because the mechanism by which gas heats in the first place is weak at low \ephot. This may be more common at earlier times, because normal star-forming galaxies at high-redshift are more compact than local galaxies at a given stellar mass \citep{Buitrago2008,Conselice2014,vanderWel2014,Mowla2019}, and \sigmaIR\ anti-correlates with \ephot. 
In addition, efficient star-formation in the thick disks of high$-z$ dusty, star-forming galaxies could contain a number of star-forming regions that each resemble the central regions of local (U)LIRGs (e.g., \citealt{Tacconi2008,Bothwell2010,Stacey2010,Brisbin2015}), in which case high $\mathrm{G/n_H}$ and low \ephot\ could be common in PDRs across a high$-z$ starburst. These conditions may play a more important role at high-redshift as the contribution of ULIRGs to the total star-formation rate density increases from $z=0$ to $z=2$ by a factor of $\sim21$ \citep{Murphy2011}. 

Recent efforts to combine mid- and far-IR measurements of gas heating and cooling at $z\sim1-2$ are largely limited to handfuls of ULIRGs with archival \textit{Spitzer}/IRS spectra \citep{Brisbin2015,McKinney2020}. Using ALMA, \cite{McKinney2020} compared the \cii\ and 6.2$\,\mu m$ PAH emission for GS IRS20, a luminous, compact starburst galaxy at $z=1.9239$ and found a low \cii/6.2$\,\mu m$ ratio at high IR surface density that could indicate a low photoelectric efficiency. This galaxy has a \lcii/$\mathrm{L_{11.3}}$\ ratio of 0.11, which would correspond to a 11.3/3.3$\,\mu m$ ratio of 21.5 if $z\sim2$ galaxies follow similar scaling relations between the PAH features and \cii\ as in local (U)LIRGs. If this is the case, then GS IRS20 would have a $3.3\,\mu m$ PAH flux of $7\times10^{-17}\,\mathrm{W\,m^{-2}}$, which can be easily detected in under 15 minutes by \textit{JWST}/MIRI. While MIRI will be excellent at measuring bright PAH lines at high$-z$, our ability to probe much of the rest-frame mid and far-IR regime at high$-z$ remains limited. Future facilities, such as the \textit{Origins Space Telescope}\footnote{\url{https://origins.ipac.caltech.edu/}}, a proposed flagship mission covering the near-far infrared with powerful imaging and spectroscopic capabilities and a large, cold telescope, would be a powerful tool for measuring the full energy budget in distant star-forming galaxies, and testing the idea that the star-formation efficiency is linked to the balance of gas cooling and heating. 

\section{Summary and Conclusion\label{sec:summary}}
We combine observations of PAH emission with measurements of the far-infrared fine-structure \cii, \oi, and \siii\ lines to infer the properties of gas heating and cooling in local, star-forming, luminous infrared galaxies. The ratio of IR cooling lines to PAH emission traces the photoelectric efficiency, \ephot, and measures the coupling between stellar radiation field and gas temperatures in photodissociation regions (PDRs). Our main conclusions are: 

\begin{enumerate}
    \item In local LIRGs, the ratio of \cii\ to PAH emission does not correlate with \lir\ or far-IR color, both of which trace a mix of the diffuse and PDR dust. 
    \item We find an anti-correlation between \cii/PAH and IR surface density (\sigmaIR) where the most compact ULIRGs have low ratios of \cii\ cooling to PAH emission compared to normal star-forming galaxies by $\sim0.5$ dex. The \cii/PAH ratio exhibits less overall scatter in GOALS than \lcii/\lir, as well as a lower magnitude of decline between low- and high-\sigmaIR\ galaxies. 
    \item \siii/PAH and \oi/PAH exhibit constant ratios up to the most compact ULIRGs, were some objects fall below the average value by a factor of $\sim5$ in both line ratios. Notably, \oi/PAH does not increase when \cii/PAH is low: enhanced cooling via the \oi\ channel is not sufficient to compensate for the deficiencies observed in the \cii\ line when considering the total cooling budget.
    \item We measure the photoelectric efficiency as (\lcii$\allowbreak+$\loi$+$\lsiii)/\lpah, which is a factor of $\sim3$ lower in galaxies with \logSigmaIR$\,>10.7$, which also have high ratios of UV radiation field strength to neutral hydrogen density (G/n$_\mathrm{H}$). Compact ULIRGs have low photoelectric efficiencies and more extreme ISM conditions, indicating a link between the large-scale energy density of a starburst and the gas cooling and heating properties on PDR scales.
    \item LIRGs with low photoelectric efficiencies have high ratios of $11.3$ to $3.3\,\mu m$ PAH emission, a tracer of the mean PAH grain size. We estimate typical grain sizes of $\mathrm{N_C}\sim60-150$ C atoms per grain in LIRGs, and find that the PAHs are predominantly ionized. Large, ionized grains produce both less and weaker photo-electrons, which may contribute to the low photoelectric efficiencies in the most compact ULIRGs if small grains are preferentially destroyed. Spectral signatures of grain emission can be used to understand the role played by dust in regulating the star-formation of galaxies. 
\end{enumerate}

The photoelectric efficiency may be key for regulating the evolution of the ISM, and can influence the overall star-formation efficiency by mediating the coupling between stellar radiation fields and gas temperatures. The trends between \ephot, \sigmaIR, and G/n$_\mathrm{H}$ reflect vigorous, compact star-formation where dusty and young PDRs exhibit less efficient gas heating. Low photoelectric efficiencies may be common in the high-redshift Universe where compact star-formation is ubiquitous, and may also contribute to changes in the star-formation efficiency. The link between the efficiency of star-formation and the cooling/heating balance will be further tested with \textit{JWST} and ALMA, but ultimately a large space-based IR telescope like \textit{Origins} is needed to measure the mid and far-infrared emission and track the full energy budget in star-forming galaxies over a significant fraction of cosmic time.
\newline

\footnotesize We thank the referee for the thoughtful comments and recommendations.~J.M. thanks S.Linden for the constructive feedback on the paper.~J.M. was supported by the IPAC Visiting Graduate Fellowship. A.S.E. and Y.S. were supported by NSF grant AST 1816838. H.I. acknowledges support from JSPS KAKENHI Grant Number JP19K23462. This work is based on observations made with the \textit{Herschel Space Observatory}, a European Space Agency (ESA) Cornerstone Mission with science instruments provided by European-led Principal Investigator consortia and significant participation from NASA. The \textit{Spitzer Space Telescope} is operated by the Jet Propulsion Laboratory, California Institute of Technology, under NASA contract 1407. This research is based on observations with AKARI, a JAXA project with the participation of ESA. This research has made use of the NASA/IPAC Extragalactic Database (NED), which is operated by the Jet Propulsion Laboratory, California Institute of Technology, under contract with the National Aeronautics and Space Administration, and of NASA’s Astrophysics Data System (ADS) abstract service. This research has made use of the NASA/IPAC Infrared Science Archive, which is operated by the Jet Propulsion Lab- oratory, California Institute of Technology, under contract with the National Aeronautics Space Administration. 

\clearpage
\bibliography{references.bib}

\end{document}